# Experimental benchmark data for Monte Carlo simulated radiation effects of gold nanoparticles. Part II: Comparison of measured and simulated electron spectra from gold nanofoils

Jorge Borbinha[1], Liset de la Fuente Rosales[2,3], Philine Hepperle[2,4], Heidi Nettelbeck[2,5], Woon Yong Baek[2], Salvatore Di Maria[1], Hans Rabus[2]

[1] Centro de Ciências e Tecnologias Nucleares, Instituto Superior Técnico, Bobadela, Portugal
[2] Physikalisch-Technische Bundesanstalt, Braunschweig and Berlin, Germany
[3] Present address: IBA Dosimetry GmbH, Schwarzenbruck, Germany
[4] Leibniz University Hannover, Institute of Radioecology and Radiation Protection, Hannover, Germany
[5] Present address: Australian National University, Acton, ACT, Australia

E-mail: hans.rabus@ptb.de

**Abstract**

Electron emission spectra of a thin gold foil after photon interaction were measured over the energy range between 50 eV and 9500 eV to provide reference data for Monte Carlo radiation-transport simulations. Experiments were performed with the HAXPES spectrometer at the PETRA III high-brilliance beamline P22 at DESY (Hamburg, Germany) for photon energies just below and above each of the gold L-edges, that is, at 11.9 keV, 12.0 keV, 13.7 keV, 13.8 keV, 14.3 keV, and 14.4 keV. The data were analyzed to obtain the absolute values of the particle radiance of the emitted electrons per incident photon flux. Simulations of the experiment were performed using the Penelope and Geant4 Monte Carlo radiation-transport codes. Comparison of the measured and simulated results shows good qualitative agreement. On an absolute scale, the experiments tend to produce higher electron radiance values at the lower photon energies studied as well as at the higher photon energies for electron energies below the energy of the Au $L_3$ photoelectron. This is attributed to the linear polarization of the photon beam in the experiments, something which is not considered in the simulation codes.

## 1. Introduction

The present work is the second part of a study aimed at providing experimental benchmark data for Monte Carlo radiation-transport simulations of the radiation effects of gold nanoparticles (AuNPs). AuNPs have an excellent biocompatibility and are intensively studied for use in radiotherapy [1–6] because in vitro and in vivo assays have shown an increase in the biological effectiveness of ionizing radiation when AuNPs are present during irradiation [5–9]. The increase in biological effectiveness is much greater than the increase in average absorbed dose expected given the greater photoabsorption of gold compared to tissue material.

This phenomenon is generally attributed to a local dose enhancement due to low-energy electrons from Auger cascades following core-shell ionizations of gold atoms [10–12]. These electrons lead to an increased energy deposition in a range of several 100 nm from the AuNP [10–15]. It has to date not been possible to measure this local energy deposition.

Numerical simulations with radiation transport Monte Carlo (MC) codes and done to determine this local dose enhancement often deliver a wide range of results between different studies [16,17]. In a recent code intercomparison exercise [18,19], the most pronounced differences between simulations were found in the energy spectra of the emitted electrons. These discrepancies persisted in part even after correcting for deviations of the simulation setups from the exercise definition [20].

At the time the exercise was conducted, most codes used in the simulations did not contain cross section data evaluated for low-energy electrons. Therefore, most exercise participants had to either use the energy cut-offs implied by the range of energies in the evaluated electron data library (EEDL) [21], often referred to as the Livermore database, or use empirical extrapolations to lower energies. Only one code used a newly developed cross section dataset for electron interactions in gold at energies down to the ionization threshold, derived from theoretical models [22]. In the meantime, cross sections for low-energy electron transport in gold have also become publicly available in the Geant4-DNA code system [23–29].

The present project was started with the aim of providing experimental benchmark data for Monte Carlo simulations of the radiation effects of AuNPs and thus indirectly for the electron cross section data implemented in the codes. The motivation and background of the study, the experimental procedures and data analysis methodology, as well as the results for the AuNP samples, were described in the first part of the paper [30]. In this second part of the paper, comparisons of the measured data for gold (Au) foils with Monte Carlo simulations using the Penelope code and Geant4-DNA are presented. The gold foil sample was included in the study because it was expected to have a better signal-to-background ratio. It was also expected that the corresponding simulations would be less demanding than for the AuNP samples. In the third part of the paper, the measured data on AuNPs and Au foil will be used for benchmarking the "radial" code [31]. A comparison of measured and simulated results on AuNP samples and a will be presented in the fourth part of the paper.

## 2. Materials and Methods

### 2.1 Experimental setup

The sample preparation and experimental procedure were described in detail in [30] and are only briefly reported here. The measurements were performed at the PETRA III undulator beamline P22 at DESY (Hamburg, Germany). The experimental setup is schematically shown in Fig. 1: A silicon double-crystal monochromator disperses the radiation emitted from an undulator, thus providing quasi-monochromatic synchrotron radiation with photon energies between 2.4 keV and 15 keV at a relative bandwidth of around $1.25 \times 10^{-4}$ and a photon flux on the order of $10^{13}$ s$^{-1}$. The beamline is equipped with the high-resolution Hard X-ray Photoelectron Spectroscopy (HAXPES) instrument, which can measure electrons with energies up to about 10 keV [32]. The investigated samples were mounted on a 5-axis manipulator with three translational and two rotational degrees of freedom [32].

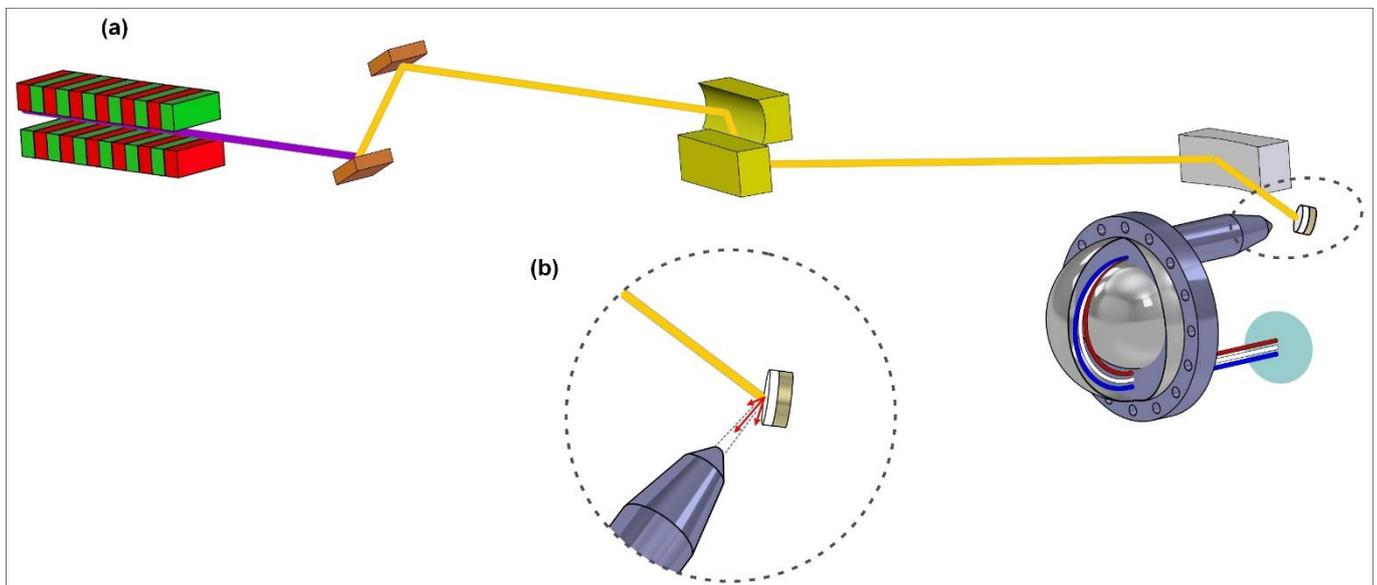

Fig. 1: (a) Schematic representation of the DESY P22 beamline with the HAXPES spectrometer. The leftmost element is the undulator, which emits a beam of broadband synchrotron radiation (blue line) which is spectrally filtered by the double-crystal monochromator. The resulting narrowband photon beam (orange line) is then focused with the mirrors (color) to the measurement position on the sample. The emitted electrons are detected with the hemispherical mirror analyzer. (b) shows a close-up image of the region around the sample indicated by the dashed line in (a). The photon beam (orange line) hits the sample surface at a grazing angle of incidence of 15°. The electrons emitted from the sample (arrows) are detected within the acceptance angle (boundaries indicated by dashed lines) of the spectrometer lens.

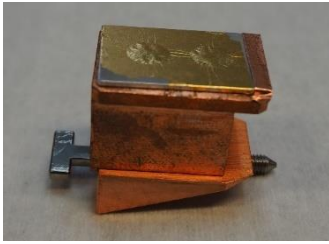

Fig. 2: Photograph of the gold-foil sample used in the measurements. The 100 nm thick self-supporting gold foil was attached to an aluminum support with two 5 mm apertures that was mounted on a DESY wedge-shaped copper sample holder.

The sample relevant to this work was a 100 nm thick gold foil attached to a 1 mm thick rectangular aluminum support measuring 22 mm × 13 mm and having two 5 mm holes. This support was manufactured at PTB (Braunschweig, Germany) and attached with double-sided carbon tape to a standard wedge-shaped copper (Cu) sample holder from DESY (Fig. 2).

Electron emission spectra from the sample were measured for energies between 50 eV and 9500 eV in steps of 1 eV. The measurements were made at 15° grazing incidence of the photon beam for six photon energies slightly below and above each of the gold L-edges, that is, 11.9 keV, 12.0 keV, 13.7 keV, 13.8 keV, 14.3 keV, and 14.4 keV. The rationale for using these energies was the relevance of the Au L-shell Auger electrons for local dose enhancement in the first few micrometers around a AuNP [30]. A 15° grazing-incidence angle was chosen to achieve a larger irradiated area (and thus to a better signal-to-noise ratio).

The measured spectra were normalized to photon flux (measured with a monitor of calibrated response) and corrected for the energy dependent transmission of the spectrometer to obtain the particle radiance of electrons per incident photon, , that is, the average number of electrons per energy interval, solid angle, and surface area. The data analysis procedure was described in detail in the first part of the paper [30].

## 2.2 Monte Carlo simulations

The main simulations of emitted electron spectra were performed for all photon energies used in the experiments with the 2018 release of the Monte Carlo radiation transport code Penelope [33]. Electron emission of a gold foil by 14.4 keV photons was also simulated using Geant4 [34–36]. To obtain additional information to facilitate the comparison of the simulation results with the experimental data, further simulations were performed using Penelope 2006.

The use of Penelope and Geant4 for this study was motivated by the fact that the results initially obtained with the two codes showed the largest differences in a recent code comparison for gold nanoparticles [20]. In addition, Geant4 was the first general radiation-transport toolkit with a low-energy extension that allows for track structure simulations [23–26], with cross sections for low-energy electron transport in gold recently added [27–29]. Penelope has also been shown to have the capability to simulate trace structures [37]. The reason for performing the complementary studies with Penelope 2006 was that a ready-to-use code with the changes described in Section 2.2.3 was available from a previous study.

### 2.2.1 Penelope 2018 simulations

The simulations with Penelope 2018 were performed using the PenEasy main code. Energy cuts of 50 eV were applied for both photon and electron transport, and the four simulation parameters $C_1$, $C_2$, $W_{cc}$ and $W_{cr}$ were set equal to 0 to force a detailed instead of a condensed-history simulation. ($C_1$ and $C_2$ are the maximum average angular deflection and the fractional energy loss, respectively, per condensed-history step. $W_{cc}$ is the lower energy cut for inelastic collisions and $W_{cr}$ is the lower energy production limit for bremsstrahlung photons.)

The geometrical setup is shown schematically in Fig. 3. The gold foil was modelled as a cuboid with dimensions of 1 cm × 1 cm × 100 nm (green line). The photon source (red line) was defined as a cuboid of 10 µm height and a square cross section with 110 µm side length. The source was located at 1 cm from the sample and emitted a parallel beam (yellow area) along the direction of its shortest dimension. The photon beam passed a vacuum-filled "direction detector" (not shown in the figure). Vacuum (implemented as hydrogen at a mass density of $10^{-16}$ g/cm³) was also the medium used for the photon source and the detector. The detector (blue rectangle) was a cylinder with a height of 100 nm and a radius of 0.158384 cm (corresponding to a half aperture angle of 9°), located at 1 cm distance from the sample in a direction perpendicular to the incident photon beam. The history of the particles was terminated when they entered the detector.

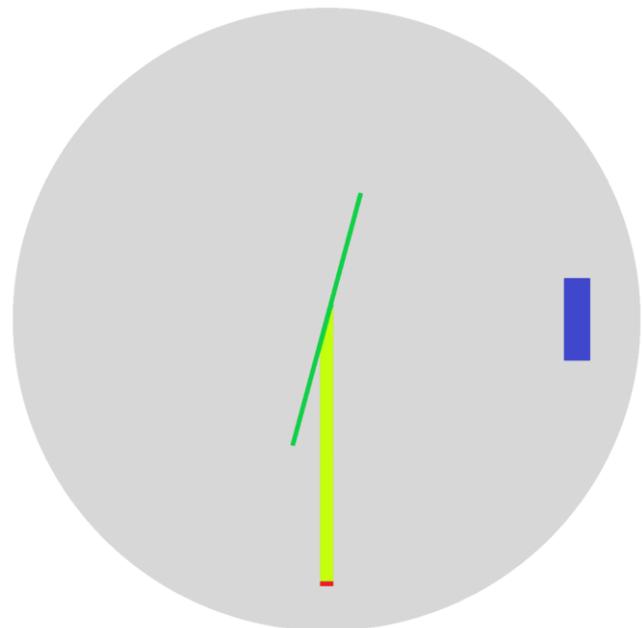

Fig. 3: Schematic cross-sectional view of the geometrical setup of the simulations with Penelope 2018 (not to scale). The gray area represents the world volume, the red line at the bottom represents the photon source, the yellow area represents the photon beam, the green line represents the gold foil and the blue area represents the detector. Except for the gold foil, all components of the geometry were filled with vacuum.

The simulations were performed for monoenergetic photons of the six energies used in the experiments in two different runs. In one run, the simulation continued until the relative standard deviation averaged over all energy bins was under 1.15 %. This corresponded to a maximum number of about $2.83 \times 10^{10}$ primary particle histories per simulation. In the second (independent) run, the simulations were stopped after a CPU time of 5 days, during which between $1 \times 10^{11}$ and $2 \times 10^{11}$ primary particle histories had been processed, depending on the photon energy. The output of a simulation from the first run was the frequency density distribution of electrons per primary particle for electron energies in the range from 50 eV to 9500 eV, divided into 1 eV energy bins. In the second run, electron energies up to the photon energy were recorded, also in energy bins of 1 eV. The two data sets were finally merged in the overlap region of the electron energies used in the experiments to improve the statistics.

### 2.2.2 Geant4-DNA simulations

The simulation was based on Geant4 version 10.7.1 [23–26] using the cross sections for track structure simulation in gold, which were kindly provided by Sakata et al. [27,28] for use in this project before they became publicly available in Geant4-DNA with version v11. In the simulations, the range cut was set to 0.1 nm for all particle types, and production cut-off energies of 50 eV were used.

The simulation setup is shown schematically in Fig. 4. The world volume was a cube of 12 cm side length, the gold foil was modeled as a cuboid of 17 mm × 22 mm × 100 nm and the detector as a cuboid of dimension 3 mm × 3 mm × 1 mm. In the simulation setup, the aluminum support of the gold foil and the copper sample holder were also included in the geometry with the dimensions used in the experimental setup.

The center of the detector was located 44 mm from the intersection of the photon beam with the sample. The radiation source was located 31 mm from this intersection and emitted photons of energy 14.4 keV in a parallel beam with a circular cross section and a radius of 0.306 µm. The direction of the photon beam had a grazing incidence angle of 15° with respect to the gold foil surface, and the short dimension of the detector was in a direction perpendicular to the photon beam.

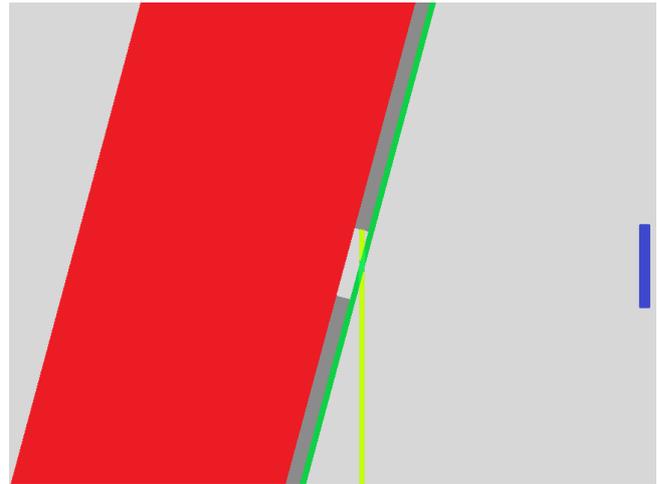

Fig. 4: Schematic cross-sectional view of a portion of the simulation setup used with Geant4-DNA (not to scale). The light-gray areas represents the vacuum-filled regions of the world volume, the red area represents the copper sample holder, the dark-gray area represents the aluminum support, the gree line represents the gold foil, the yellow line represents the incident photon beam, and the blue area represents the detector.

When an electron entered the detector, its energy was recorded, and the particle history was terminated. The simulations were performed in batches and multithreaded mode for a total of $8 \times 10^9$ primary particles. The output of the simulations was a list of the phase space data of the registered electrons, which were evaluated using ROOT and a custom-build GDL [38] script to determine the frequency density per primary photon at 1 eV intervals between 0 eV and 14400 eV electron energy.

### 2.2.3 Penelope 2006 simulations

The simulations with Penelope 2006 were performed with a variant of the "penmain" program included in the code distribution, which was modified in several ways. First, rather than using an external photon beam, the primary photons were generated within the gold foil and forced to interact at their initial position. This position was uniformly sampled within a parallelepiped defined by the intersection of a square cylinder with a cuboid. The side length of the square cylinder was hard coded as 100 µm (which roughly corresponds to the photon beam in the experiments). The cuboid had dimensions of 2 cm × 2 cm × 100 nm (corresponding to the thickness of the gold foil). The axis of the cylinder passed through the center of the cuboid and was at an angle of 15° to the planes delimiting the cuboid in the direction of its smallest dimension. The initial direction of the photons was along this axis, and the simulation geometry consisted only of the cuboid.

The second change in the program was to record the initial position when phase-space data of electrons directly produced by a photon interaction was retrieved from the secondary particle stack. This information was inherited by daughter particles of these electrons and their descendants.

The history of electrons crossing the surface of the foil was terminated after determining whether their continued trajectory would intersect the circular entrance aperture of the electron spectrometer. Two different distances between this aperture and the sample were considered, namely the focal length of the electron spectrometer (54 mm) and the actual distance used in the experimental setup (46 mm). The size of the circle was such as to give a half opening angle of 9° at the focal length.

The results of these simulations were frequency density distributions of electron energy in the energy range between 50 eV and 14.4 keV in 1 eV intervals for the two distances between the sample and the aperture. Additional output files were created listing the final energy for each detected electron along with the initial energy and depth below the front surface of the gold layer. These simulations were performed for all six photon energies. In a later modification, the simulation output (produced for 14.4 keV photons only) also included the lateral offset (as seen from the spectrometer) of the point at which the electron left the foil from its initial position. The purpose of this additional output was to assess the size of the secondary electron source and to confirm that the different geometries used in the simulations could be corrected by scaling the solid angle covered by the spectrometer entrance aperture.

### 2.3 Methodology of comparison between experiment and simulation

The quantity determined in the simulations is the frequency density distribution, meaning, the number of electrons per primary photon and per electron energy interval emitted from the sample within the solid angle subtended by the detector. The output of the experiments is the particle radiance of electrons per incident photon, meaning the number of electrons per energy interval, solid angle, and surface area. To make the two outcomes comparable on an absolute scale, the simulation results have to be converted into an estimate of particle radiance.

A simple approach that gives the correct dimensions would be to divide the frequency density by the solid angle of the detector and the surface area covered by the photon beam in the simulations. However, it must be taken into account that the number of counted electrons in the simulations depends only slightly on the beam cross section, as long as this is negligibly small compared to the dimensions of the aperture representing the detector. The reason for this is that the variation of the radiance with the angle $\theta$ between emission direction and the axis of the spectrometer is expected to be small when the angle of acceptance is not too large.

For isotropic emission, the radiance would vary with $\cos\theta$, and for a half opening angle of 9°, the variation between the center and the circumference of the aperture is about 1.2 %. For a p-wave of photoelectrons originating from an s shell, the dependence is $\cos^2\theta$ and leads to a maximum variation of about 2.4 %. In the experiments, the photon beam had a lateral extension (in one direction) of ± 200 µm. This elongation corresponds to a range of angle of ± 1.15° for the Penelope 2018 simulations where the detector was closest to the irradiated surface. Thus, the difference in the number of counted electrons is expected to be in the sub-percent range when a different photon beam size is used, as in the Geant4 simulations.

A quantitative comparison must therefore consider for the simulation results the same value for photon beam area $A_b$ as in the experiments as well as the actual solid angle subtended by the detector in the simulations. Following this rationale, the frequency density $f_s(E_e)$ from the simulations was converted into the simulated particle radiance, $d^3\varepsilon_s/dA\,dE\,d\Omega$, according to Eq. (1).

$$\frac{d^3\varepsilon_s(E_e)}{dA\,dE\,d\Omega} = f_s(E_e) \times \frac{d_a{}^2}{A_b A_a} \qquad (1)$$

where $A_a$ and $d_a$ are the area of the aperture and its distance from the irradiated sample surface in the simulations.

### 3. Results and Discussion

#### 3.1 Measured electron spectra

The measurement results obtained for the Au foil sample are shown in Fig. 5, where the three panels on the left show the dependence on the kinetic energy of the electrons and the three panels on the right show the dependence on the binding energy. The three rows of plot panels correspond to measurements slightly below and above the $L_3$, $L_2$, and $L_1$ absorption edges of gold, respectively. In Fig. 5(a), a sharp increase of the particle radiance below 7.5 keV electron energy is seen for the spectrum at 12.0 keV photon energy as compared to 11.9 keV. This increase is due to the opening of additional Auger decay channels after the creation of a $L_3$ core hole by photoabsorption. According to the Evaluated Atomic Data Library (EADL) [39], this core hole leads to the emission of an $L_3M_xM_y$ Auger electron with energies between about 5 keV and 7.5 keV with a probability of almost 50 %, and in about 17 % of the cases a non-radiative filling of this hole produces Auger electrons with energies above 7.5 keV. In addition, there is an approximate 31 % probability of deexcitations by a radiative transition producing a fluorescence photon [39].

Much smaller differences are seen after photoabsorption on the $L_2$-shell (Fig. 5(b) and (e)). According to the EADL, an $L_2$ vacancy is filled in 35 % of cases by a radiative transition and in about a third of cases by emission of an electron in the energy range between 7.5 keV and 9.5 keV, as can be seen in Fig. 5(b). For this shell, there is already a probability of about 11 % for the occurrence of Coster-Kronig (CK) transitions leading to the emission of electrons with energies below 2 keV [39].

Experimental benchmark data for Monte Carlo simulated radiation effects of gold nanoparticles. Part II: Comparison of measured and simulated electron spectra from gold nanofoils

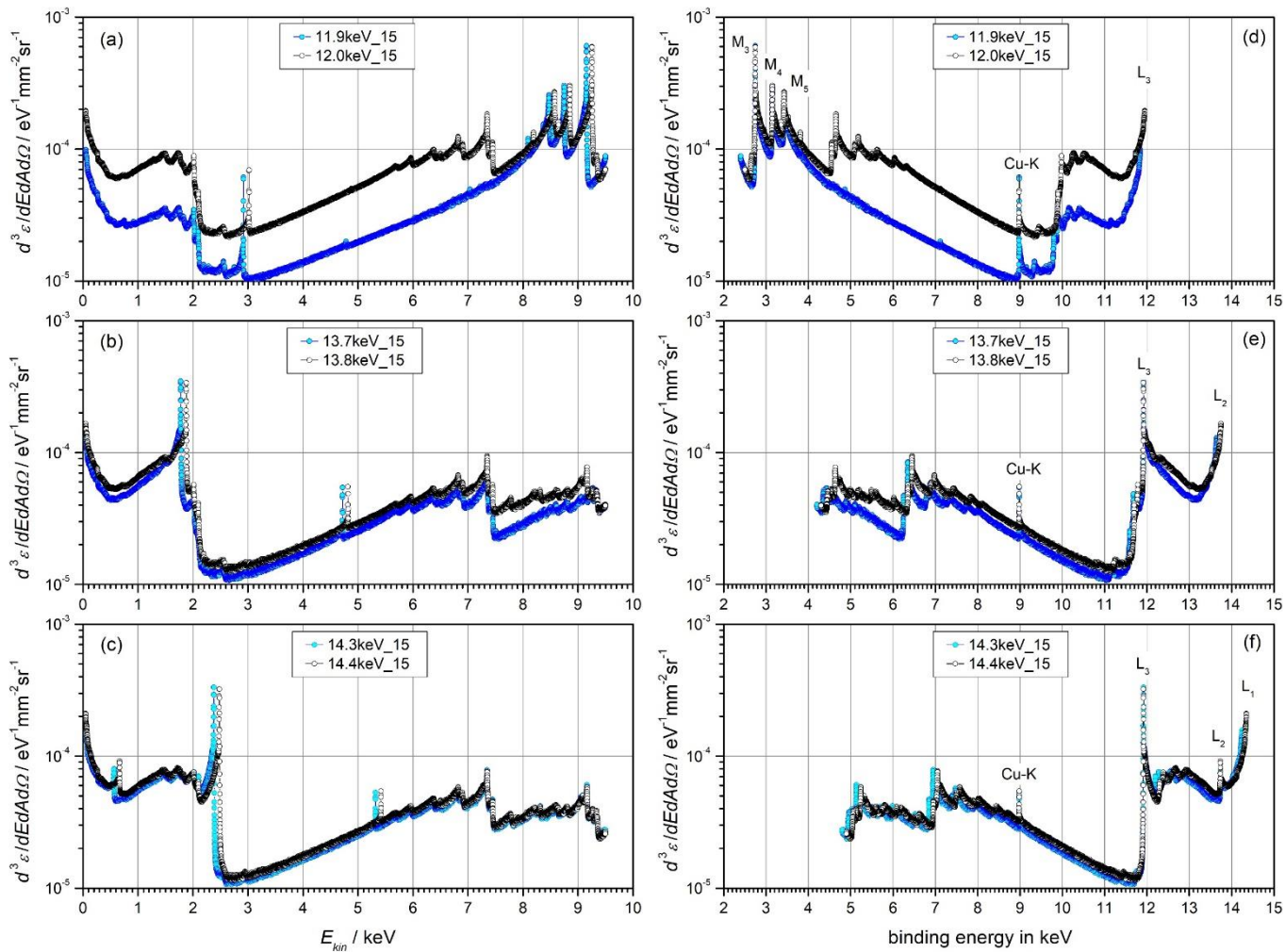

Fig. 5: Comparison of results for the particle radiance per photon flux of the 100-nm Au foil measured at an incidence angle of 15° plotted (a) to (c) against the electron kinetic energy and (d) to (f) against binding energy. In (a) and (d), the photon energies are below (11.9 keV) and above (12.0 keV) the $L_3$ absorption edge of gold, in (b) and (e) below (13.7 keV) and above (13.8 keV) the $L_2$ edge, and in (c) and (f) below (14.3 keV) and above (14.4 keV) the $L_1$ absorption edge.

When the photon energy increases from just below to just above the $L_1$ absorption edge, only small changes are seen in Fig. 5(c) and (f) with the logarithmic y-scale. For the $L_1$-shell, 73 % of the vacancies are filled by CK transitions leading to energies of the emitted electrons of below 2.5 keV, and three quarters of them even have energies below 250 eV [39]. In 7 % of cases, a fluorescence photon is emitted, and in another 7 % of cases an Auger electron with energy exceeding 9.5 keV is emitted [39].

As can further be seen in the panels on the right side of Fig. 5, the photoemission spectrum of the Au foil also contains a pronounced peak of Cu K photoelectrons. In addition, it was shown in the first part of the paper that silver (Ag) L lines can also be detected in the energy region of the Au M-shell photoelectrons [30]. The occurrence of such impurities is expected given the good alloyability of the three noble metals.

### 3.2 Comparison of measured and simulated spectra

The comparison of the measured particle radiance spectra with the particle radiance obtained from the simulation results by applying Eq. (1) is shown in Fig. 6, where the different panels correspond to the different photon energies. Larger deviations between experiment (blue circles) and simulation (yellow diamonds) can be seen in Fig. 6(a) and (b) for the lowest two photon energies (11.9 keV and 12.0 keV), where the experimental results are higher than the simulated data by about a factor of 2. In addition, the experimental results are significantly lower in the first few hundred eV of electron energy in all panels. This suggests the presence of a surface contamination layer that hinders the escape of low-energy electrons [40,41] but presumably has little effect on electrons with energies greater than 500 eV.

As the photon energy increases, the difference between measured and simulated results generally decreases, except

Experimental benchmark data for Monte Carlo simulated radiation effects of gold nanoparticles. Part II: Comparison of measured and simulated electron spectra from gold nanofoils for the very different shapes of the photoelectron and Auger electron peaks and except for the energy region below the L-shell photoabsorption peaks. Moreover, the energies of the Auger peaks show significant deviations between experiment and simulation, which is related to the fact that the energy values used in the Penelope code come from the EADL, which differ from the values found experimentally, as already observed in [30]. The reason is that the energy positions in the EADL were obtained for the case of sudden transitions, where the binding energies of the electrons are assumed to be unchanged, whereas in reality some relaxation of the excited system is expected during the short, but still non-zero, lifetime of the core vacancies.

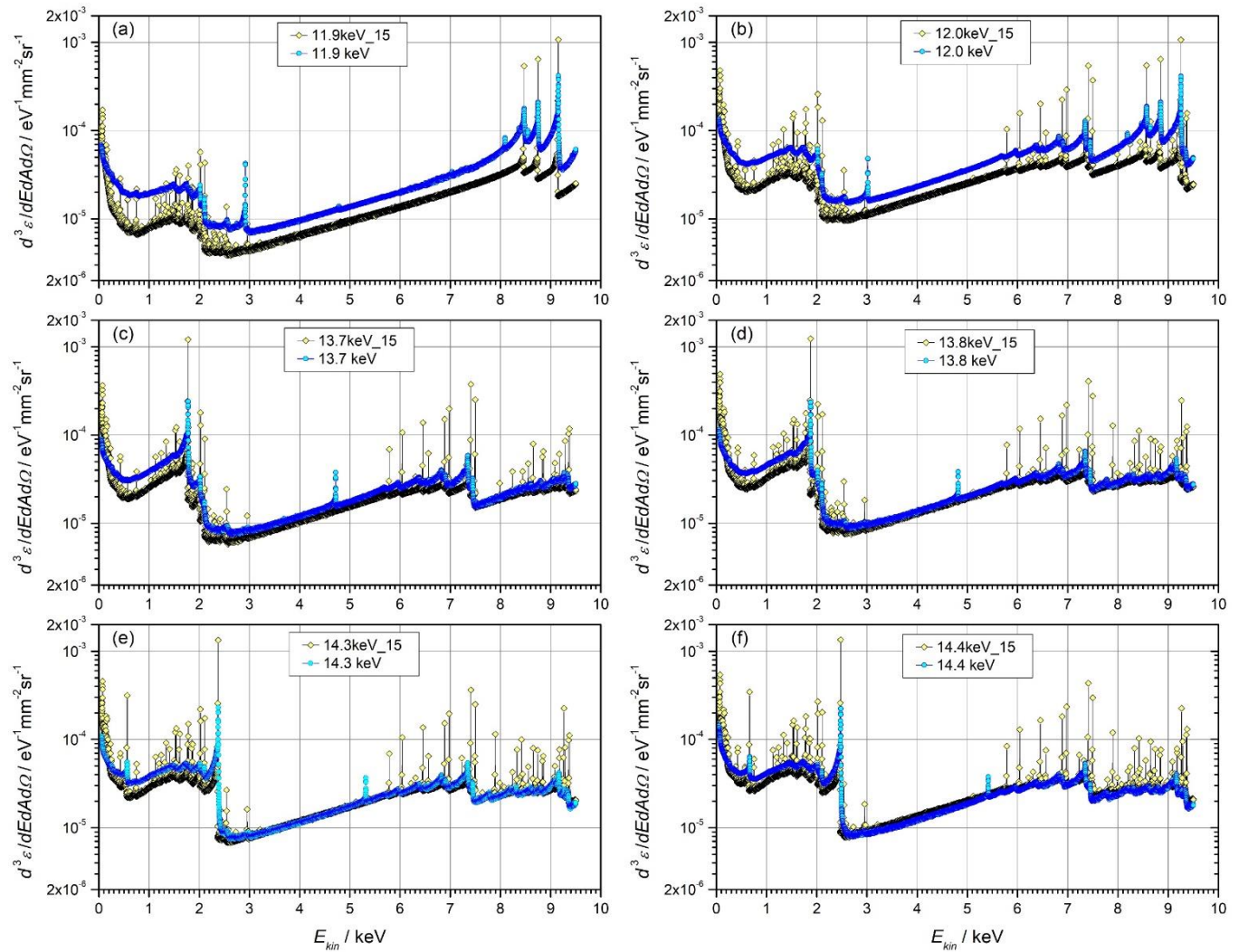

Fig. 6: Comparison of results for the particle radiance per photon flux of the 100-nm Au foil measured at an inicidence angle of 15° for photons with energies near the (a) and (b) $L_3$, (c) and (d) $L_2$, and (e) and (f) $L_1$ absorption edges of gold. The open blue circles represent the experimental data and the yellow-filled symbols represent the simulation results obtained with Penelope 2018.

The lifetime of the core vacancies is also the reason for the very different line shapes of the peaks observed in the measurements compared to those obtained in the simulations. In the simulations, an infinitely sharp energy peak is assumed for both the photoelectrons and the Auger and Coster-Kronig electrons, while in the experiments the short lifetime of the core hole leads to a significant line broadening.

However, it should be noted that the agreement between experiment and simulations at higher photon energies may be a coincidence, since the experimental results have an uncertainty as large as about 40 % [30]. The main contributions to this uncertainty come from the electron spectrometer transmission and the photon beam size, which affect the data in the same way at different photon energies. The other contributions to the experimental uncertainty total only about 6 %. The variation of the difference between experiment and simulations with photon energy is therefore far outside the experimental uncertainty.

### 3.3 Quantitative comparison

The different peak energies and line shapes make the comparison of simulation results and experimental data at each electron energy meaningless. To quantitatively compare simulated and measured data, two different approaches were followed. In the first approach, it was tested how well simulations and experiments agree when the simulation results are modified to account for broadening due to finite core-hole lifetime and the experimental resolution.

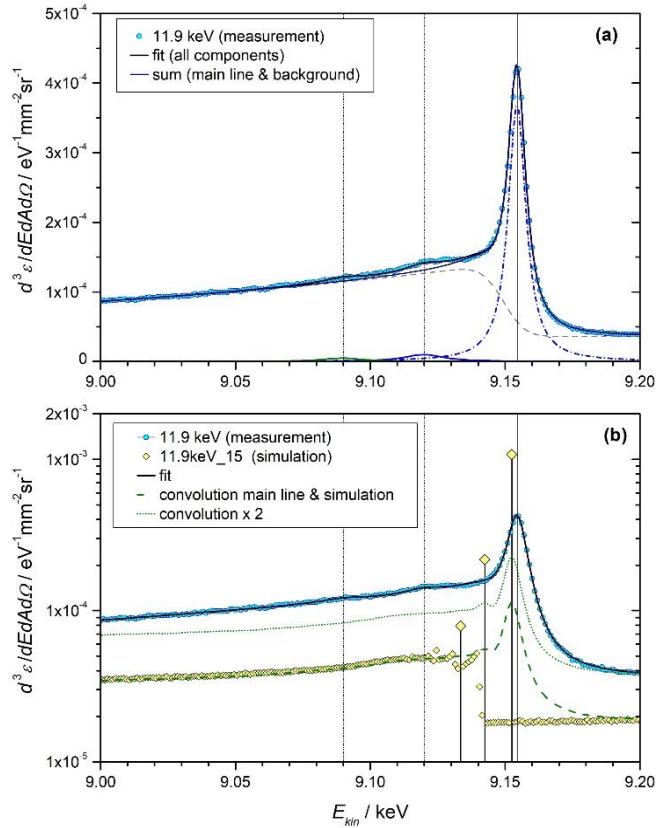

Fig. 7: Comparison of the experimental data (blue circles) in the energy range of the Au $M_3$ photoelectron line (at 11.9 keV photon energy) with (a) the results of a line fit (solid black line) and (b) with the Penelope 2018 simulation results (diamonds) and the curve obtained by convolution of the simulation results with the profile of the main line, which is given by the dot-dashed curve in (a). (For details see text.)

Fig. 7 shows an example of this approach in the energy region around the Au $M_3$ photoelectron line for 11.9 keV photon energy. Fig. 7(a) shows the experimental data together with the fit curve and the fitting components considered: a main line representing the photoelectron peak (dot-dashed); a background with (different) exponential slope below and above the photoelectron energy and a transition in the form of an error function; and two additional peaks at lower energy (energy position indicated by the vertical dot-dashed lines). The thick solid line is the sum of all components. The thin solid line is the sum of the background curve (dashed) and the main line.

The main photoelectron line has been fitted by a Voigt profile, that is, by the convolution of a Lorentz (life-time broadening) and a Gauss (instrumental broadening) curve. The full-width at half maximum (FWHM) of the Gaussian profile was set to 1 eV, which corresponds to the bandwidth of the photon beam. The FWHM of the Lorentz curve was obtained from the fit as 7.7 eV, corresponding to a life-time of about 0.5 fs. The two additional peaks at 9090 eV and 9120 eV are very broad structures and have been fitted by Lorentz curves of 18 eV FWHM. This large linewidth indicates the possible presence of several peaks. However, their small intensity, which is only slightly larger than the noise, hampers a more detailed analysis.

Fig. 7(b) shows the comparison between the experimental data and the simulation results. In this enlarged view (compared to Fig. 6(a)) it is evident that the simulation yields three pronounced peaks in this energy range (larger diamonds), which are separated by 9 eV and 10 eV, respectively. These electrons are likely to have undergone one or two collisions producing plasmons, which are modeled with a fixed energy in the simulations.

The long-dashed line in Fig. 7(b) is the convolution of the simulation results with the profile of the main line obtained when fitting the experimental data (dot-dashed line in Fig. 7(a)). It is evident that taking account of the broadening increases the similarity of the line shapes in the simulation to those observed experimentally. However, the difference of a factor of about 2 on average remains. The short-dashed line is the convoluted curve multiplied by a factor of 2, which brings the simulation results into agreement at higher energies. In the energy range of the main peak and at lower energies, there are still discrepancies of the order of a factor of 2. This could be related to the fact that linearly polarized radiation was used in the experiments, which was not taken into account in the simulations.

The second approach to a quantitative analysis is illustrated in Fig. 8, which shows how several global indicators vary with photon energy. For 14.4 keV, two data points are shown for each indicator, one from Penelope 2018 and one from Geant4 (open symbols). The first two are the mean (black squares) and the median (red circles) of the point-by-point ratio of simulated and measured particle radiance. The error bars refer to the data points of the mean and indicate the standard deviation of the ratios for all electron energies. The median is expected to be a more robust estimator. However, only minor differences between the median and mean can be seen.

The other data points represent the ratios between simulation and experiment for the integral of the particle radiance (upward triangles) and of the energy-weighted particle radiance (downward triangles) over the energy range between 50 eV and 9500 eV. (The integral was estimated by summing over all energy bins and multiplying by the bin width.) The first quantity is the number of electrons emitted per area, solid angle and incident photon; the second is the total energy transported by emitted electrons per area, solid angle and incident photon.

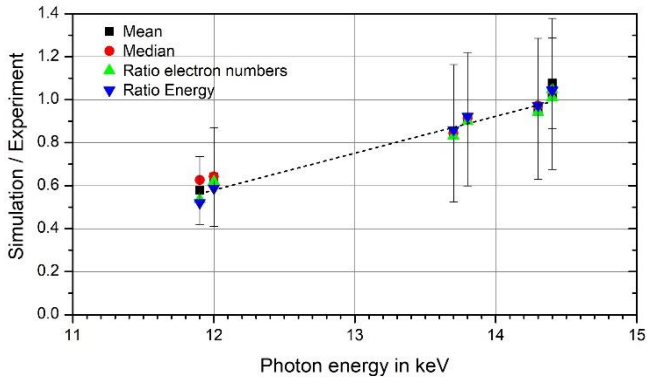

Fig. 8: Mean and median of the point-by-point ratio of the simulation results to the experimental data, and ratios of the integrated particle radiance and energy radiance from simulation to experiment. The error bars refer to the data points of the mean and indicate the standard deviation of the point-by-point ratios. The dashed black line is merely a guide to the eye.

All the data shown in Fig. 8 seem to follow the dashed line that suggests a factor of about 2 variation between the lowest and highest photon energies. This variation is smaller than the variation observed for AuNPs in [30], where the ratio of the estimated total energy transported by emitted electrons per incident photon to the photon energy was higher by a factor of 3 to 5 at 11.9 keV compared to 14.4. keV. This was explained in part by a signal contribution from a Tougaard background of electrons produced in the copper sample holder [42,43] and in part by the anisotropic electron emission resulting from the polarization of the photon beam.

From the values given in the XCOM database [44], it can be estimated that the transmission of a 100-nm gold foil at 15° grazing incidence is 95 % for 11.9 keV and 87 % at 12.0 keV photons. The latter is also the value for photons of 13.8 keV and 14.4 keV. However, it is unlikely that the measurement on the gold foil also contains a background of electrons released after photon interaction in the aluminum support or the copper sample holder. Gold is a much stronger scatterer of electrons than carbon. Therefore, a background of electrons generated in the sample holder can be expected to be at least an order of magnitude smaller in the Au foil sample than in the AuNP sample, whose carbon foil substrate is only half as thick as the gold foil.

This is supported by the results of the Geant4 simulations, which took into account the presence of the aluminum support and copper sample holder. The Geant4 simulation results are compared in Fig. 9(a) with the data from the Penelope 2018 simulation and in Fig. 9(b) with the experimental values. To account for the poor statistics of the Geant4 results, all data have been rebinned to 10 eV intervals.

The comparison shows generally good agreement of the Geant4 results with the other data. Significant discrepancies from the Penelope results can be seen in Fig. 9(a) at electron energies below 1 keV and between 2.5 keV and 3.5 keV. Discrepancies from the measurements can be seen in Fig. 9(b) between 2.5 keV and 4 keV. These latter discrepancies indicate that the Geant4 simulations predict a small Tougaard background originating from Cu K photoelectrons, which was not found in the measurements. This is presumably due to the fact that the gold foil in the experiments was not a parallel slab (Fig. 2), as assumed in the simulations. It is therefore possible that the angle of incidence to the gold surface was slightly different from 15°. Here a smaller angle would reduce the photon transmission of the foil and increase the yield of escaping electrons. In any case, a contribution of electrons from the sample holder to the experimental results can be ruled out.

In contrast, the occurrence of electrons produced in the Au foil by interactions of fluorescence photons generated in the Cu sample holder may not be negligible. The corresponding signal contribution was estimated based on the linear photon attenuation coefficients of Au and Cu from the XCOM database [44]. The fluorescence photon energies of Cu and the corresponding transition probabilities were obtained from the evaluated atomic data library (EADL) [39,45]. The photon interaction cross sections for the M shells of Au were taken from the database provided with the Penelope code [46].

Cu $K_\alpha$ photons occur with a probability of about 39 % and have energies slightly above 8 keV. Cu $K_\beta$ photons have energies slightly below 9 keV and occur with a probability of about 5 %. For the irradiation at 15° grazing incidence, the estimated ratio of the flux of Cu $K_\alpha$ and $K_\beta$ fluorescence photons to the flux of photons from the beamline was about $3\times10^{-3}$ and $4\times10^{-4}$, respectively. This estimate is based on the assumption that the fluorescence photons can produce detectable electrons over the central 10 mm² of the gold foil surface. The self-absorption of the fluorescence photons in the Cu material was taken into account.

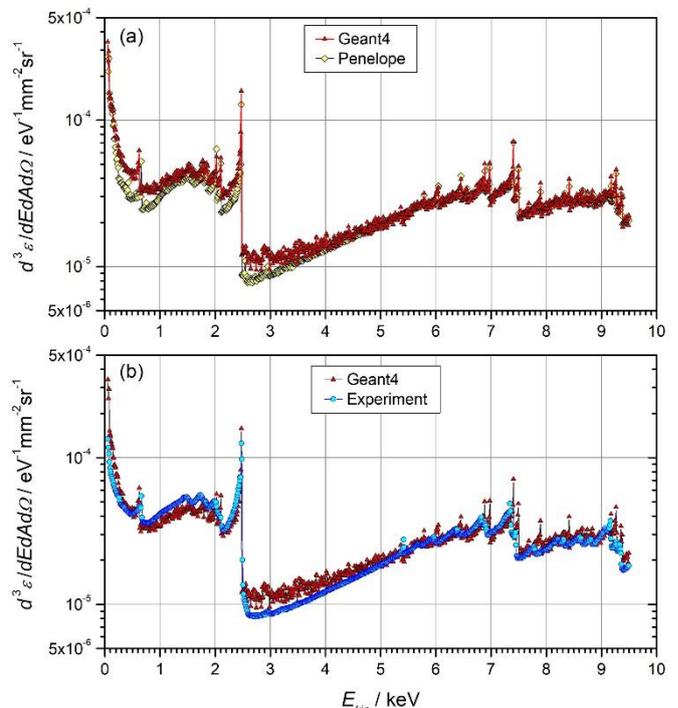

Fig. 9: Comparison of the Geant4 simulation results for 14.4 keV photon energy with the results obatined with Penelope (a) and from the experiments (b). The experimental data and those from Penelope have been rebinned to match the 10 eV bin size of the Geant4 data.

This additional photon flux leads to the production of Au M-shell photoelectrons with energies between 4.6 keV and 5.8 keV for Cu $K_\alpha$ and between 5.6 keV and 6.8 keV for Cu $K_\beta$ photons. The estimated ratio between the peak intensity of these photoelectrons and that of those produced by the incident photons is between 0.7 % and 0.9 % for the Au $M_1$ line and between 1.4 % and 2.2 % for the Au $M_5$ line. (The lower values apply to 11.9 keV photon energy, the higher ones to 14.4 keV.) These values are negligibly small and explain why the corresponding peaks are not seen in Fig. 6.

On the other hand, photon polarization could be an explanation for the variation by a factor of 2 found in Fig. 8. The significantly higher experimental values compared to the simulations occur at the higher photon energies for electron energies below the L-shell photoelectron peaks and for the two lowest photon energies in the entire electron energy range (Fig. 6). This suggests that the occurrence of these higher values is related to the fact that the experiments were performed with synchrotron radiation, which is linearly polarized.

It should be noted that this does not contradict the observation evident in Fig. 7(b) that multiplying the convolution of the simulation data with the photoelectron line by a factor of 2 brings it into agreement with the measured data at higher energies. The residual discrepancies between the short-dashed line in Fig. 7(b) and the data represented by circles occur around the peak energy and for lower electron energies. The peak corresponds to the $M_3$ photoelectron, meaning that there are at least two more photoelectron lines at energies outside the energy range covered by both experiments and simulations. These electrons also produce a background that scales with the intensity of the photoelectron peak. Therefore, polarization effects that lead to increased photoelectron peaks also mean a corresponding increase in the background originating from these electrons.

To avoid confusion, it should be noted that preliminary simulations using Penelope 2018 were performed both with and without the Stokes parameters for photon polarization specified in the input files [33]. The results of the two simulations differed only slightly, in the range of a few percent. The simulations leading to the results shown in Fig. 6 were performed with Stokes parameters for 100 % linear polarization. Inspection of the source code revealed that the Stokes parameters are used for modeling the (elastic) photon scattering, but are not considered for modelling the photoabsorption process, where the azimuthal angle of the photoelectron is sampled from a uniform distribution between 0 and $2\pi$. Therefore, the major discrepancies between measurements and simulations are attributed to polarization of the photon beam.

### 3.4 Relevance the results

The results of this study show that benchmarking MC simulations of emitted electron spectra with measured data presents challenges beyond obtaining experimental data on an absolute scale and achieving good simulation statistics. For the AuNP samples studied [30], the mass per area of gold was by about two orders of magnitude lower than that of the gold foil.

Therefore, a standard approach in the MC simulation would require at least two orders of magnitude more CPU time to achieve statistics comparable to those of the Au foil. And the statistics of the Au foil simulations were already significantly worse than in the experimental data. This is explained by the fact that the number of incident photons corresponding to an experimental spectrum was in the order of $10^{16}$. For comparison, in the simulations of the gold foil, the number of histories processed per CPU core amounted to about $10^8$ per day. As will be discussed in detail in the fourth part of the paper, sophisticated variance reduction methods need to be applied in the AuNP sample simulations.

The measured data for the gold foil are still quite useful, as simulations for this 'bulk-like' system are obviously easier to perform and can be used in studies for an initial test of the simulation setup and the appropriate choice of simulation parameters, such as the interaction data ("physics list").

A caveat concerning the experimental data is their comparatively large measurement uncertainty. However, a major part of this uncertainty comes from in the transmission of the electron spectrometer and is thus independent of the photon energy. Any better estimate of the "true" transmission value would therefore affect the data in the same way for all photon energies and all samples. This means that the difference between simulated and measured results should be the same for both the Au foil and an AuNP sample. This provides an additional benchmarking criterion.

As pointed out in [30], the scatter of the results of simulations of GNP radiation effects reported in the literature is much larger than the uncertainties of the experimental results [16,18,19]. Thus, the fact that the experimental dataset produced in this study is only for six photon energies and is subject to large uncertainties does not seem to hinder its use as a benchmark. A table of the experimental results is therefore provided as Supplement 1 to this article.

### 4. Conclusions

This work presented a quantitative comparison between measured and simulated energy spectra of electrons emitted from a gold foil under irradiation with monochromatic photons of energies in the range of the Au L-shells. To our knowledge, this is the first study of this kind. The use of absolute quantities rather than (potentially rebinned) relative spectral distributions has yielded several interesting insights.

With respect to the absolute magnitude of the particle radiance of the electrons emitted in the solid angle subtended by the electron spectrometer and of the energy transported by them, a pronounced dependence on the photon energy was found for the difference between experimental and simulated results. This photon-energy-dependent discrepancy can be attributed to the fact that the measurements were performed with linearly polarized radiation, while photon polarization was not considered in the modelling of the photoelectric absorption process in the simulations.

Furthermore, it was shown that convolving the simulation results with a function representing lifetime and instrumental broadening yields a spectral shape comparable to that observed in the measurements. It could be an interesting (and presumably easy to implement) extension of radiation-transport codes to include lifetime broadening in the modelling of photoabsorption and incoherent scattering on inner shells and the ensuing de-excitation of heavy elements. Modifying the photoelectric absorption models to account for photon polarization also appears to be a straightforward improvement to the codes.

It is understood that these details are not expected to result in significant changes in the assessment of absorbed dose in sufficiently large scoring volumes, which is often the primary purpose of radiation transport codes. However, they may be as relevant for track structure simulations as are cross sections for low-energy electron transport. Although potential use cases generally do not involve synchrotron radiation, only synchrotron radiation sources provide sufficiently high photon flux to perform benchmark experiments of the kind presented here. Since such benchmark studies are an essential part of quality assurance for computational dosimetry, it would be beneficial if the codes were able to simulate such experiments while taking account of actual experimental conditions including photon beam polarization.

## 5. Acknowledgements

This work was funded by the German Research Foundation (DFG) under grant number 386872118. DESY (Hamburg, Germany), a member of the Helmholtz Association HGF, is acknowledged for the provision of experimental facilities. Parts of this research were carried out at PETRA III, and the authors would like to thank Christoph Schlueter, Andrei Gloskovskii and Patrick Lömker for their assistance in using beamline P22 with the HAXPES setup. Beamtime was allocated for proposal I-20200068. Andreas Pausewang is acknowledged for his support in the preparation of the beamtime and during the experiments. Florian Burger and Gert Lindner are acknowledged for their assistance in running some of the simulations on the PTB high-performance computing cluster, Leo Thomas for cross-checking the Geant4 simulations and Miriam Schwarze for useful discussions related to these simulations, and Ronald Dunham for linguistic proofreading of the manuscript. The authors also thank Dousatsu Sakata for early access to the low-energy cross sections for electron scattering in gold that allowed the Geant4-DNA simulation to be performed prior to making the cross sections available to the public in release v11.

## 6. Author contributions:

**HR:** Methodology, Validation, Data Curation, Formal Analysis, Supervision, Visualization, Writing - Original Draft, Writing - Review & Editing; **JB, LFR, PH, LT:** Investigation, Writing - Review & Editing; **HN:** Investigation, Supervision, Writing - Review & Editing; **SDM, WYB:** Supervision, Writing - Review & Editing